\documentclass[12pt]{iopart}
\usepackage{iopams}

\usepackage{graphics}
\usepackage{graphicx}
\usepackage{epsfig}

\begin{document}

\letter{Chaos suppression in the large size limit for long-range systems}

\author{Marie-Christine Firpo\dag\ and Stefano Ruffo\dag \ddag}

\address{\dag\ Dipartimento di Energetica ''Sergio Stecco'',\\ 
Universit\`{a} degli Studi di Firenze, Via Santa Marta, 3, I-50139 Firenze, Italy}
\address{\ddag INFM and INFN, Firenze}


\begin{abstract}
We consider the class of long-range Hamiltonian systems first introduced by Anteneodo and 
Tsallis and called the $\alpha$-XY model. This involves $N$ classical rotators on a 
$d$-dimensional periodic lattice
interacting all to all with an attractive coupling whose strength decays as $r^{-\alpha}$, 
$r$ being the distances between sites. Using a recent geometrical approach, we estimate 
for any $d$-dimensional lattice the scaling of the largest Lyapunov exponent (LLE)
with $N$ as a function of $\alpha$ in the large energy regime where rotators behave almost freely. We find that the LLE vanishes as
$N^{-\kappa}$, with $\kappa=1/3$ for $0 \le \alpha/d \le 1/2$ and $\kappa=2/3(1-\alpha/d)$
for $1/2 \le \alpha/d < 1$.
These analytical results present a nice agreement with
numerical results obtained by Campa \textit{et al.}, including deviations at small $N$.
\end{abstract}


\submitto{\JPA}

\maketitle

It is well known that systems interacting via long-range interactions may exhibit 
pathological thermodynamical as well as dynamical
behaviours. This issue has been very much debated recently. In particular, for systems 
governed by sufficiently long-range interactions decaying as
$r^{-\alpha}$, with $\alpha \le d$ the Euclidean space dimension and 
$r$ the interparticle distance, the Hamiltonian comes 
out to be \textit{non-extensive}, that is the energy per particle
diverges in the thermodynamic limit $N\rightarrow \infty$. In order to study 
issues related to the links between non-extensivity and
long-range interactions, Anteneodo and Tsallis introduced in 
reference~\cite{Anteneodo98} a generalization of the Hamiltonian mean-field 
(HMF)~\cite{Antoni95} ferromagnetic-like X-Y model in the form 
\begin{equation}
H=\frac{1}{2}\sum\limits_{i=1}^{N}L_{i}^{2}+\frac{1}{2\tilde{N}}%
\sum\limits_{i,j=1,i\neq j}^{N}\frac{1-\cos \left( \theta _{i}-\theta
_{j}\right) }{r_{ij}^{\alpha }}=K\left( \mathbf{L}\right) +V\left( \mathbf{%
\theta }\right)  \label{Ham}
\end{equation}
with $d \geq \alpha \geq 0$. This is an Hamiltonian model of rotators placed at the sites of a 
$d$-dimensional lattice with indices $i$, $j$ and $r_{ij}$ is the shortest
distance between them with periodic boundary conditions, so that the
interaction between rotators $i$ and $j$ decays as the inverse of their
distance to the power $\alpha $. Here, the $\tilde{N}$ rescaling function
is introduced in order to get an {\it extensive} (i.e. of order $N$) 
potential $V$. This trick was not adopted in reference \cite{Anteneodo98},
but then the authors had to conveniently rescale thermodynamical
potentials by $\tilde{N}$. In the large $N$ limit \cite{Tsallis95} 
\begin{equation}
\label{Ntilde}
\tilde{N}\equiv 1+d\int\limits_{1}^{N^{1/d}}r^{d-1-\alpha }dr\sim 
\cases{N^{1-\alpha /d}&if $\alpha \neq d $\\
\ln \left( N\right) &if $\alpha =d$\\}  
\end{equation}
As remarked in reference \cite{Barre} this does not entail energy {\it additivity}.
In the one-dimensional case, Tamarit and Anteneodo \cite{Tamarit} have shown 
numerically that the canonical caloric and magnetization curves for model (\ref{Ham}) could
be derived from the curves of the HMF model (recovered for $\alpha =0$, $d=1$), 
and this result has been later derived analytically 
by Campa, Giansanti and Moroni~\cite{Campa2000}.

Recently, the stochasticity exhibited by system (\ref{Ham}) has been
investigated numerically through the computation of the largest Lyapunov
exponent (LLE) as a function of the energy density 
$\varepsilon =H/N$ \cite{Anteneodo98,Campapreprint,Giansantipreprint}. In the phase where particles behave almost freely, i.e. for $\varepsilon $ above
the critical energy density $\varepsilon _{c}$, where the system exhibits a
second order phase transition, the LLE has been found to
scale with the number of rotators as $N^{-\kappa }$ where $\kappa $ is a
so-called ``universal'' function  of the ratio $\alpha /d$ with no 
dependence on the energy $\varepsilon $
nor on $d$. In the past, the LLE had been studied for the HMF 
model~\cite{Latora98,Latora99,Firpo98}.

For Hamiltonian models having a large number of degrees of freedom like (\ref
{Ham}) and using a geometric reformulation of the dynamics that associates
trajectories to geodesics, Casetti, Pettini and coworkers \cite
{Cas95,Cas96,Cas2000} have proposed an expression of the LLE in terms of the
ensemble-averaged curvature and fluctuations of curvature of the mechanical
manifold associated to the Hamiltonian. This geometrical approach has proved
to give very accurate estimates in a large number of Hamiltonian physical
systems for which chaos mainly originates from parametric instability~\cite{VanKampen}, that
is systems for which the curvature is mainly positive but fluctuating. The
method was originally applied to derive estimates of the LLE in the
thermodynamic limit $N\rightarrow \infty $. Later on, this approach enabled
the analytical computation \cite{Firpo98} of the LLE for the HMF model as a
function of $\varepsilon $. In the homogeneous phase 
($\varepsilon>\varepsilon _{c}$) in which the LLE, denoted hereafter $\lambda _{1}$, vanishes 
in the limit $N\rightarrow \infty $, it was shown that keeping the
leading order in $N$ in the ensemble averages of geometric quantities
enabled to derive the scaling of $\lambda _{1}$ as a function of $N$. In 
this way, the LLE was predicted to scale as $N^{-1/3}$ for 
$\varepsilon>\varepsilon _{c}$. The aim of this calculation was to relate the behaviour
of the dynamical indicator $\lambda _{1}$ to the occurrence of a second order
phase transition at $\varepsilon _{c}$ in the system.

In this Letter we wish to apply the same approach to the generalized model 
(\ref{Ham}) in order to give an analytical prediction for the exponent 
$\kappa $. We shall use the derivation of the canonical thermodynamics for
the system (\ref{Ham}) presented in reference~\cite{Campa2000}.

Let us first recall the expression for the LLE derived from the geometric 
approach~\cite {Cas95,Cas96,Cas2000}. The effective curvature felt by a geodesic is modeled
as a Gaussian stochastic process whose mean is the average Ricci curvature and
the variance its fluctuations. Under the ergodic hypothesis, these quantities
may be replaced by their microcanonical ensemble-averages, denoted
respectively $\kappa _{0}$ for the curvature and $\sigma _{\kappa }^{2}$ for
its fluctuations. Then~\cite{Cas95}
\begin{equation}
\lambda _{1}=\frac{1}{2}\left( \Lambda -\frac{4\kappa _{0}}{3\Lambda }\right)~,
\label{expla}
\end{equation}
where 
\begin{equation}
\Lambda =\left( \sigma _{\kappa }^{2}\tau +\sqrt
{\left( 4\kappa_{0}/3\right) ^{3}+\sigma _{\kappa }^{4}\tau ^{2}}\right) ^{1/3}~,
\label{bigL}
\end{equation}
and where $\tau$ is a time scale for the stochastic process estimated as 
\begin{equation}
\tau =\frac{1}{2}\left[ \frac{2\sqrt{\kappa _{0}+\sigma _{\kappa }}}{\pi }+%
\frac{\sigma _{\kappa }}{\sqrt{\kappa _{0}}}\right] ^{-1}.  \label{tau}
\end{equation}
Using Eisenhart metric, the microcanonical average $\kappa _{0}$ of the mean
Ricci curvature $k_{R}$ reads $\kappa _{0}=\left\langle k_{R}\left( \mathbf{%
\theta }\right) \right\rangle _{\mu }$ where 
\begin{equation}
k_{R}\left( \mathbf{\theta }\right) \equiv \frac{K_{R}\left( \mathbf{\theta }%
\right) }{N-1}=\frac{1}{N-1}\sum_{i=1}^{N}\frac{\partial ^{2}V}{\partial
\theta _{i}^{2}}  \label{meanRC}
\end{equation}
and 
\begin{equation}
\sigma _{\kappa }^{2}\equiv \left\langle \delta ^{2}k_{R}\right\rangle _{\mu
}=\frac{1}{N}\left\langle \left( K_{R}-\left\langle K_{R}\right\rangle _{\mu
}\right) ^{2}\right\rangle _{\mu }.  \label{fluct}
\end{equation}
As explained in references \cite{Firpo98,Cas96,Cas2000} we shall assume the
equivalence of microcanonical and canonical ensembles, leading to identical
values of both ensemble-averages of observables in the limit $N\rightarrow
\infty $, but to different values of their fluctuations with the formula 
\cite{Leb67}
\begin{equation}
\left\langle \delta ^{2}f\right\rangle _{\mu }=\left\langle \delta
^{2}f\right\rangle _{c}+\left( \frac{\partial \left\langle \varepsilon
\right\rangle _{c}}{\partial \beta }\right) ^{-1}\left[ \frac{\partial
\left\langle f\right\rangle _{c}}{\partial \beta }\right] ^{2},
\label{LebVer}
\end{equation}
where $\beta \equiv 1/T$ and $k_{B}=1$. This assumption is fully justified 
by the recent achievement that ensemble inequivalence for averages in long-range 
systems is to be expected only close to first order phase
transitions~\cite{Gross,Barre,Cohen}. Therefore, the first step is to express 
the canonical thermodynamics of 
system (\ref{Ham}) following reference \cite{Campa2000}. The constraint $i\neq j$
in the potential can be removed for free by defining $r_{ii}^{-\alpha }=b$,
that is for the time being an arbitrary constant. Then the symmetric distance
matrix $R_{ij}^{\prime}=r_{ij}^{-\alpha }$ may be diagonalized, which enables to use the Hubbard-Stratonovitch transform to evaluate the potential
part $Z_{Vc}$ of the canonical partition function. Using then the
saddle-point method, one obtains from equations (12) and (20) of reference \cite
{Campa2000} in the long-range case ($\alpha \leq d$) and for 
$\varepsilon >\varepsilon _{c}$ (zero magnetization phase) 
\begin{equation}
\ln Z_{Vc}=-\frac{\beta }{2\tilde{N}}\sum\limits_{j=1}^{N}r_{ij}^{-\alpha} 
-\frac{1}{2}\sum_{n=1}^{N}\ln \left[ 1-\beta 
\frac{\lambda _{n}}{\tilde{N}}\right]  \label{lnZcv}
\end{equation}
The eigenvalues $\lambda_n$ of the $R^{\prime}$ matrix can be easily
derived following reference \cite{Moroni}.

Let us first consider the $d=1$ case for the sake of simplicity. We remind that 
\begin{equation}
\label{defmatrixRp}
R_{ij}^{\prime }=
\cases{r_{ij}^{-\alpha }&if $i\neq j$ \\
b&if $i=j$\\}  
\end{equation}
where 
\begin{equation}
r_{ij}=\min_{l\in \mathbb{Z}}\left| i-j+lN\right|.  \label{defrij}
\end{equation}
Therefore $R_{ij}^{\prime }=R^{\prime }(i-j)=R^{\prime }(m)$ where $%
R^{\prime }$ is a $N$-periodic function. This periodicity of the lattice
enables to diagonalize $R^{\prime }$ in Fourier space. Its Fourier transform
is 
\[
\tilde{R}^{\prime }(n)=\sum\limits_{m=1}^{N}\exp (-i\frac{2\pi }{N}%
nm)R^{\prime }(m) 
\]
with the inversion formula 
\[
R^{\prime }(j)=\frac{1}{N}\sum\limits_{k=1}^{N}\exp (-i\frac{2\pi }{N}jk)%
\tilde{R}^{\prime }(k). 
\]
Then it can be easily shown that 
$R^{\prime}(i-j)=\sum\nolimits_{k=1}^{N}u_{ik}^{\dagger }\lambda _{l}u_{jk}$ 
where $u_{jk}:=N^{-1/2}\exp (-i\frac{2\pi }{N}jk)$ is an element of the unitary
matrix of eigenvectors with the following expression for the eigenvalues 
($1\leq k\leq N$) 
\begin{equation}
\lambda _{k}=\sum\limits_{m=1}^{N}\exp (-i\frac{2\pi }{N}km)R^{\prime }(m).
\label{defeigenv}
\end{equation}
For any $d$-dimensional lattice, one would get the generalized expression for 
$\lambda _{\mathbf{n}}=\lambda \left( n_{1},\ldots ,n_{d}\right) $ with $%
1\leq n_{1},\ldots ,n_{d}\leq N^{1/d}$ as
\begin{equation}
\lambda _{\mathbf{n}}=\sum\limits_{m_{1}=1}^{N^{1/d}}\ldots
\sum\limits_{m_{d}=1}^{N^{1/d}}\exp \left( -i\frac{2\pi }{N^{1/d}}%
\sum\limits_{i,j=1}^{d}n_{i}m_{j}\right)R^{\prime }(\mathbf{m}).   \label{vpdimd}
\end{equation}
Coming back to the $d=1$ case, we shall take $N$ even in the following and put $N=2p$. As $R^{\prime }$ is
an even function, this implies, for $1\leq k\leq N=2p$, 
\begin{equation}
\lambda _{k}=b+\tilde{\lambda}_{k}  \label{defdetev}
\end{equation}
with 
\begin{equation}
\tilde{\lambda}_{k}:=\frac{\left( -1\right) ^{k}}{p^{\alpha }}%
+2\sum\limits_{m=1}^{p-1}\frac{\cos \left( \pi km/p\right) }{m^{\alpha }}.
\label{defdetevtilde}
\end{equation}
$\tilde{\lambda}_{p}$ is the smallest of the $\tilde{\lambda}_{k}$'s and is
negative. In order to get a fully positive spectrum that enables to apply
the Hubbard-Stratonovitch transform, we can now shift the spectrum by 
\textit{fixing} 
\begin{equation}
b:=-\tilde{\lambda}_{p}=-\frac{\left( -1\right) ^{p}}{p^{\alpha }}
-2\sum\limits_{m=1}^{p-1}\frac{\left( -1\right) ^{m}}{m^{\alpha }}.
\label{fixb}
\end{equation}
$\tilde{N}$ is then defined as the maximal eigenvalue 
\begin{equation}
\tilde{N}=\lambda _{2p}=b+\frac{1}{p^{\alpha }}+2\sum\limits_{m=1}^{p-1}%
\frac{1}{m^{\alpha }}  \label{defNtildefull}
\end{equation}
and it can be easily checked that 
\begin{equation}
\sum\limits_{k=1}^{N=2p}\tilde{\lambda}_{k}=0.  \label{sommenulle}
\end{equation}

We can now go on with the derivation of the
LLE. Using expression (\ref{meanRC}), one gets 
\begin{equation}
k_{R}\left( \mathbf{\theta }\right) =\frac{1}{N-1}\frac{1}{\tilde{N}}%
\sum\limits_{i\neq j}^{N}\frac{\cos \left( \theta _{i}-\theta _{j}\right) }{%
r_{ij}^{\alpha }}=\frac{1}{N-1}\left[ \frac{1}{\tilde{N}}\sum\limits_{i\neq
j}^{N}r_{ij}^{-\alpha }-2V\left( \mathbf{\theta }\right) \right]~.
\label{courb1}
\end{equation}
Thus 
\begin{equation}
\left\langle k_{R}\left( \mathbf{\theta }\right) \right\rangle _{c}=\frac{1}{%
N-1}\left[ \frac{1}{\tilde{N}}\sum\limits_{i\neq j}^{N}r_{ij}^{-\alpha }+2%
\frac{\partial \ln Z_{Vc}}{\partial \beta }\right]~.  \label{courb2}
\end{equation}
That is 
\begin{equation}
\left\langle k_{R}\left( \mathbf{\theta }\right) \right\rangle _{c}=\frac{1}{%
N-1}\left[ -\frac{N}{\tilde{N}}b+\sum_{n=1}^{N}\frac{\lambda _{n}}{\tilde{N}%
-\beta \lambda _{n}}\right]  \label{courbure}
\end{equation}
Let us now consider the fluctuations of the curvature. Their canonical average
is 
\begin{equation}
\left\langle \delta ^{2}k_{R}\right\rangle _{c}=\frac{4}{N-1}\frac{\partial
^{2}\ln Z_{Vc}}{\partial \beta ^{2}}=\frac{2}{N-1}\sum_{n=1}^{N}\left( \frac{%
\lambda _{n}}{\tilde{N}-\beta \lambda _{n}}\right) ^{2}~,  \label{fluc1}
\end{equation}
while the corrective term (\ref{LebVer}), needed to get the fluctuations in
the microcanonical ensemble, is
\begin{equation}
\left( \frac{\partial \left\langle \varepsilon \right\rangle _{c}}{\partial
\beta }\right) ^{-1}\left[ \frac{\partial \left\langle k_{R}\right\rangle
_{c}}{\partial \beta }\right] ^{2}=-\frac{1}{2}\beta ^{2}\left\langle \delta
^{2}k_{R}\right\rangle _{c}^{2}.  \label{corfluc}
\end{equation}
Actually, the energy density above the critical point is intensive and equal
to $1/(2\beta )$ (up to a constant term). And one obtains 
\begin{equation}
\sigma _{\kappa }=\left( 1-\frac{1}{2}\beta ^{2}\left\langle \delta
^{2}k_{R}\right\rangle _{c}\right) ^{1/2}\left\langle \delta
^{2}k_{R}\right\rangle _{c}^{1/2}.  \label{flucmicro}
\end{equation}

We shall focus on the high
temperature regime and derive the scaling of $\lambda _{1}$ with $N$ under
both the limits of large $N$ and small $\beta $. We will show later the validity of this scaling of $\lambda _{1}$ with $N$
in a wider range of $\beta $. The results will be
compared with the numerical results of references \cite
{Campapreprint,Giansantipreprint} for which $\beta =1/9$. We shall also
compute a numerical estimate of the scaling of $\lambda _{1}$ for $\beta
=1/9 $ using directly equations (\ref{expla}-\ref{bigL}) with the full expressions (\ref
{courbure}-\ref{fluc1}-\ref{flucmicro}).

Using (\ref{courbure}), one gets 
\begin{equation}
\left\langle k_{R}\left( \mathbf{\theta }\right) \right\rangle _{c}\sim
_{\beta \rightarrow 0^{+}}\frac{\beta }{\left( N-1\right) \tilde{N}^{2}}%
\sum_{n=1}^{N}\lambda _{n}^{2}  \label{equivcourb}
\end{equation}
where $\sum\nolimits_{n=1}^{N}\lambda _{n}^{2}=\sum\nolimits_{n=1}^{N}\tilde{%
\lambda}_{n}^{2}+Nb^{2}$. Using (\ref{defdetevtilde}), one obtains $%
\sum\nolimits_{n=1}^{N}\tilde{\lambda}_{n}^{2}=Np^{-2\alpha
}+2N\sum\nolimits_{m=1}^{p-1}m^{-2\alpha }$. If $2\alpha \neq 1$, the sum of
the $\tilde{\lambda}_{n}^{2}$'s is of the order $N^{2-2\alpha }$. Therefore,
using (\ref{Ntilde}) and (\ref{defNtildefull}), the scaling of $\kappa _{0}$
with $N$ depends on the value of $\alpha $ as 
\begin{equation}
\label{courbscal}
\kappa _{0}\propto _{\beta \rightarrow 0^{+}}
\cases{\beta N^{-1}&for $0\leq \alpha <1/2$ \\
\beta N^{-2+2\alpha }&for $1/2\leq \alpha <1$ \\
\beta  \left( \ln N\right) ^{-2}&for $\alpha =1.$\\}
\end{equation}
Let us now estimate the order of the fluctuations. From (\ref{fluc1}), (\ref
{flucmicro}) and in the limit of vanishing $\beta $, $\sigma _{\kappa }^{2}$
is equivalent to 
$2\left( N-1\right) ^{-1}\tilde{N}^{-2}\sum\nolimits_{n=1}^{N}\lambda _{n}^{2}$, so that 
\begin{equation}
\label{fluctscal}
\sigma _{\kappa }^{2}\propto _{\beta \rightarrow 0^{+}}
\cases{N^{-1}&for $0\leq \alpha <1/2$ \\
N^{-2+2\alpha }&for $1/2\leq \alpha <1$ \\
\left( \ln N\right) ^{-2}&for $\alpha =1.$\\}
\end{equation}
At this stage we can check that the results presented in reference \cite{Firpo98}
are effectively recovered for the HMF case in the high energy regime where
the curvature and fluctuations of the curvature were predicted to be of order 
$N^{-1}$. Now, considering the time scale estimated as (\ref{tau}), one
obtains $\sqrt{\kappa _{0}+\sigma _{\kappa }}\sim \sqrt{\sigma _{\kappa }}$
that vanishes in the limit of large $N$ while $\sigma _{\kappa }/\sqrt{%
\kappa _{0}}\sim \beta ^{-1/2}=$ $\mathcal{O}(1)$. This last estimate is
interpreted in references \cite{Cas96,Cas2000} as the relevant timescale when the
fluctuations are of the same order as the curvature and does not require the
positivity of the curvature, which is effectively the case here. It comes
out then that $\tau $ is of order one. Putting this together with estimates 
(\ref{courbscal}) and (\ref{fluctscal}) in equations (\ref{bigL}) and 
(\ref{expla}) leads to the expression of the LLE, in the limit of large $N$ and at the
leading order in $N$, as 
\begin{equation}
\label{scalfin}
\lambda _{1}\left( N,\beta \ll 1\right) \propto _{\beta \rightarrow
0^{+}}\beta ^{1/6}\times 
\cases{N^{-1/3}&for $0\leq \alpha <1/2$ \\
N^{-2/3+2\alpha /3}&for $1/2\leq \alpha <1$ \\
\left( \ln N\right) ^{-2/3}&for $\alpha =1.$\\} 
\end{equation}
where one can also check the integrability of the model in the large temperature $\beta \rightarrow
0^{+}$ limit for all $N$ \cite{correction}. If one develops the curvature at higher orders in $\beta $, one obtains a
power series of the type $\beta \sum\nolimits_{n=1}^{N}\left(
\lambda _{n}/\tilde{N}\right) ^{2}+\beta ^{2}\sum\nolimits_{n=1}^{N}\left(
\lambda _{n}/\tilde{N}\right) ^{3}+\ldots $ where it can be checked that all
the sums of the successive powers of $\lambda _{n}/\tilde{N}$ are equivalent
in the large $N$ limit. Similar expansions are obtained for the
fluctuations of curvature. In the large $N$ limit, one can thus conclude
that equation (\ref{scalfin}) is valid for small $\beta $, and not only in the $%
\beta \rightarrow 0^{+}$ limit. Formula (\ref{scalfin}) is hence still valid after replacing $\beta ^{1/6}$ with a function of $%
\beta $ whose power series development in $\beta$ is in principle calculable. Moreover, coming back to the general $d$-dimensional case, the
previous procedure can be followed replacing (\ref{defeigenv}) by (\ref{vpdimd}). The sum
(\ref{equivcourb}) involves then
\[
\sum\limits_{n=1}^{N}\tilde{\lambda}_{n}^{2}\sim
N\sum\limits_{m_{1}=1}^{N^{1/d}}\ldots \sum\limits_{m_{d}=1}^{N^{1/d}}\left(
m_{1}^{2}+\ldots +m_{d}^{2}\right) ^{-\alpha }\sim NN^{1-2\alpha /d}
\]
if $2\alpha /d\neq 1$. Therefore we can replace $\alpha $ by $\alpha /d$ in
all the previous scalings. This analysis predicts then the universal exponent $\kappa $ as 
\begin{equation}
\label{univge}
\kappa =\kappa \left( \alpha/d \right) =
\cases{1/3&if $0\leq \alpha/d <1/2$ \\
\frac{2}{3}(1-\alpha/d) &if $1/2\leq \alpha/d <1.$\\}
\end{equation}
Hence $\kappa$ is a function of $\alpha /d$ which is equal to $1/3$ for $\alpha/d =0$,
consistently with numerical and analytical results obtained for the HMF
model \cite{Latora98,Latora99,Firpo98}. It vanishes in the limit $\alpha
/d\rightarrow 1$ consistently with the intensivity of the LLE for
short-range ($\alpha \geq d$) potentials predicted from thermodynamics 
\cite{Campa2000}. Moreover we predict that for $\alpha=d$ the LLE should scale as $
\left( \ln N\right) ^{-2/3}$ in the limit of large $N$ and energy density. This
agrees with the intuitive statement made in reference \cite{Campapreprint} that
the LLE should scale as some power of $1/\ln N$ for $\alpha =d$.

\begin{figure}[tbp]

  \centerline{
  \psfig{figure=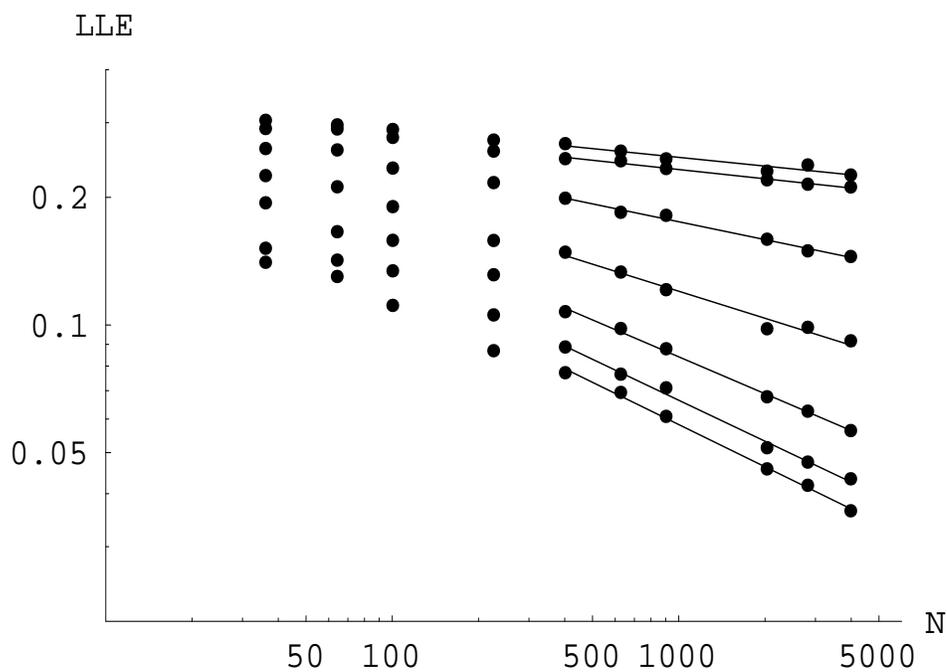,width=13cm,height=9cm}
  }
  \vskip5mm
\caption{Loglog plot of the LLE as a function of $N$ for different values of 
$\alpha$ with $d=2$. The data have been communicated by the authors of reference~\cite{Campapreprint}. 
At each value of $\alpha/d$, the best linear fit of the logarithm of the LLE as a function of the logarithm of $N$ is plotted for $N$ between 400 and 4000.  The 
values of $\alpha/d$ are, from top to bottom, 1, 0.95, 0.8, 0.6, 0.4, 0.2 and 0.}
\label{fig001}
\end{figure}

\begin{figure}[tbp]
  \centerline{
  \psfig{figure=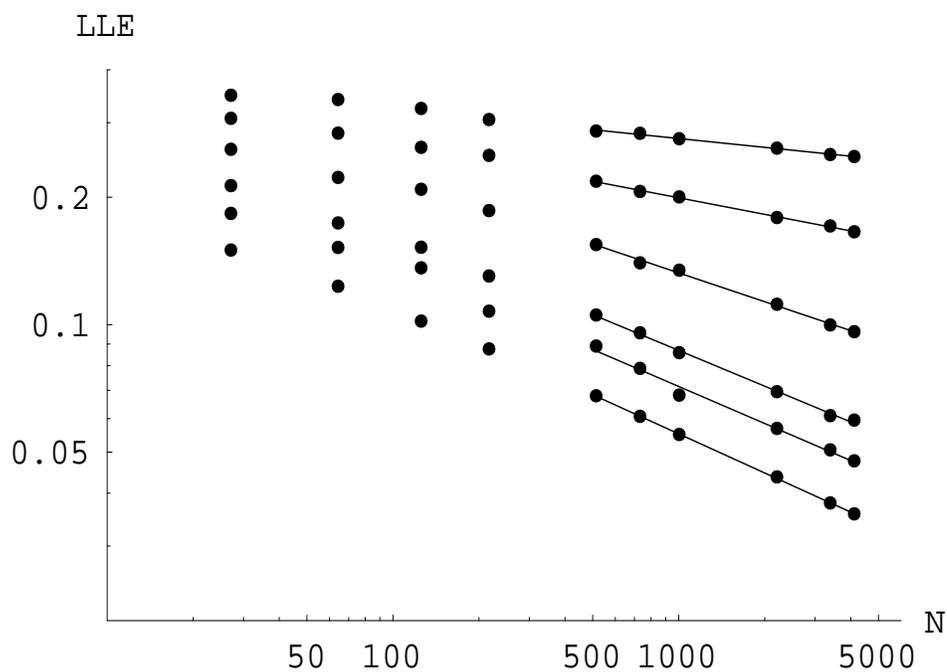,width=13cm,height=9cm}
  }
  \vskip5mm
\caption{ Same plot as in figure~\ref{fig001} with $d=3$. The values of 
$\alpha/d$ are, from top to bottom, 0.95, 0.8, 0.6, 0.4, 0.2 and 0.}
\label{fig002}
\end{figure}

\begin{figure}[tbp]
  \centerline{
  \psfig{figure=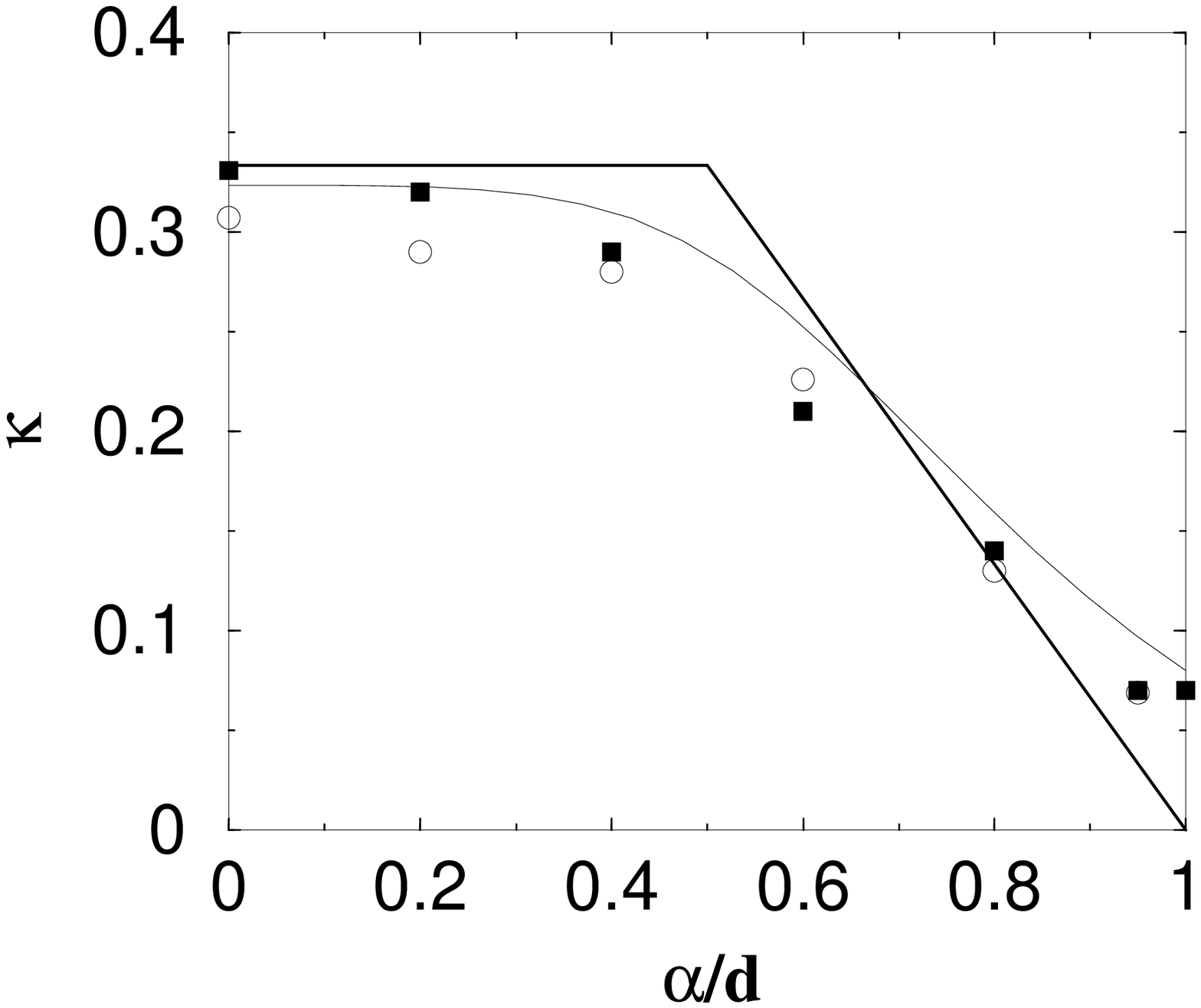,width=13cm,height=9cm}
  }
  \vskip5mm
\caption{Plot of the exponent $\kappa$ as a function of $\alpha/d$. The bold line corresponds 
to the analytical prediction obtained in the infinite $N$ limit. The thinner line gives the 
exponent obtained by fitting the logarithm of the analytical prediction of the LLE as a function
of the logarithm of $N$ for $N$ between 500 and 4000. This is to be compared with the values 
of $\kappa$ as a function of $\alpha/d$ deduced from the fits shown in figures~\ref{fig001} 
(with filled squares) and~\ref{fig002} (with empty circles).}
\label{fig003}
\end{figure}

We shall now compare these estimates
with the numerical results presented in reference \cite{Anteneodo98} for $d=1$
and in references \cite{Campapreprint,Giansantipreprint} for $d=1,2$ and 3. In these 
papers, the LLE has been computed numerically for values of
$N$ ranging from 5 to 1000 for $d=1$ and from 36 to 3969 for $d=2,3$. In 
references~\cite{Anteneodo98,Campapreprint,Giansantipreprint}, the curves giving the 
LLE as a function of $N$ in log-log scale were
fitted by the functional form $aN^{-\kappa} + bN^{-\kappa-c}$ with $c=1$. We do 
not agree that this form is the correct second order approximation for 
any value of $\alpha/d$ between 0 and 1, nor that $c$ should be equal to 1. Moreover, 
due to obvious numerical constraints, the values of $N$
used are not sufficiently large to discard additional finite-$N$ effects. For the values used 
in references \cite{Campapreprint,Giansantipreprint} for $d=2$ and $d=3$ we instead performed 
a linear fit retaining only the highest values of $N$, with $N$ 
between 500 and 4000, as shown in figures \ref{fig001} and \ref{fig002} and got the 
numerical values of $\kappa$ plotted in figure \ref{fig003}. 
On the other hand, using (\ref{expla}-\ref{bigL}-\ref{tau}) together with 
(\ref{courbure}-\ref{fluc1}-\ref{flucmicro}), we can compute the LLE for different values
of $\alpha$ and $N$ and obtain the theoretical curves associated to the numerical ones 
in figures \ref{fig001} and \ref{fig002}. In log-log
scale, we have derived the best linear fits from the theoretical curves giving the 
LLE as a function of $N$ for different values of $\alpha$. 
It should be mentioned that the results so far obtained for $\kappa$ benefit from a 
better statistics for the largest values of $N$ compared to the numerical 
results of references~\cite{Campapreprint,Giansantipreprint}, which explains that the theoretical
curve be slightly above numerical points in figure~\ref{fig003}.
Nevertheles we observe in this figure a good agreement between the theoretical and 
numerical finite-$N$ derivation of $\kappa$ as a function of
$\alpha/d$. The agreement is worse for $d=3$, compared with the $d=2$ case; this may be
related to more important finite $N$ effects as $d$ increases for a given value of $N$. This supports the validity of our theoretical predictions. 
Consequently, we suggest that the so-called 'universal curves' published
in references \cite{Anteneodo98} and \cite{Campapreprint,Giansantipreprint} are 
plagued by finite-$N$ effects. In the asymptotic limit in $N$, we
claim that the correct universal scaling of the LLE with $N$ is given by (\ref{univge}). 
This means that there is a sharp change in the dynamical behaviour of the model for $\alpha/d=1/2$. 
For $0 \le \alpha/d \le 1/2$, the suppression of chaos scales like in the HMF model, which provides 
then a universal law in this range, with a universal exponent $\kappa=1/3$.

\ack We acknowledge useful discussions with A. Giansanti, F. Leyvraz and C. Tsallis. 
We thank C. Anteneodo and the authors of references \cite
{Campapreprint,Giansantipreprint} for communicating us their numerical data. 
We also thank the organizer, A. Rapisarda, and the
participants of the HMF meeting held in Catania in September 2000 where
issues relevant for this work were discussed. MCF thanks European
Commission for support through a Marie Curie individual fellowship Contract
No HPMFCT-2000-00596. This work is also part of the contract COFIN00 on 
\textit{Chaos and localization in classical and quantum mechanics}.


\clearpage

\begin{thebibliography}{99}
\bibitem{correction} We have to remark that
there is a mistake in equation (23) of reference \cite{Firpo98} for the HMF
($\alpha=0$) model. It should be read 
$\lambda_{1} \sim 4^{1/3}c^{1/2}\beta^{1/6}(2-\beta c)^{-1/2}N^{-1/3}$,
which is actually of order $\beta^{1/6}$ in the small $\beta$ limit.
\end{thebibliography}
\end{document}